\documentclass[aip, amsmath,amssymb, reprint,]{revtex4-1}

\usepackage{graphicx}
\usepackage{dcolumn}
\usepackage{bm}

\usepackage[utf8]{inputenc}
\usepackage[T1]{fontenc}
\usepackage{mathptmx}
\usepackage{etoolbox}
\usepackage{xcolor}
\usepackage{soul}

\makeatletter
\def\@email#1#2{%
 \endgroup
 \patchcmd{\titleblock@produce}
  {\frontmatter@RRAPformat}
  {\frontmatter@RRAPformat{\produce@RRAP{*#1\href{mailto:#2}{#2}}}\frontmatter@RRAPformat}
  {}{}
}%
\makeatother

\begin{document}

\author{S. Battisti}
 \affiliation{NEST, Istituto Nanoscienze-CNR and Scuola Normale Superiore, Pisa, Italy}
\author{J. Koch}
\affiliation{Department of Physics, University of Konstanz, 78457 Konstanz, Germany}
\author{A. Paghi}
\affiliation{NEST, Istituto Nanoscienze-CNR and Scuola Normale Superiore, Pisa, Italy}
\author{L. Ruf}
\affiliation{Department of Physics, University of Konstanz, 78457 Konstanz, Germany}
\author{A. Gulian}
 \affiliation{Advanced Physics Laboratory, Institute for Quantum Studies, Chapman University, Burtonsville, MD 20866, USA}
\author{S. Teknowijoyo}
 \affiliation{Advanced Physics Laboratory, Institute for Quantum Studies, Chapman University, Burtonsville, MD 20866, USA}
\author{C. Cirillo}
\affiliation{CNR-Spin, c/o Università degli Studi di Salerno, I-84084 Fisciano (Sa), Italy}
\author{Z. Makhdoumi Kakhaki}
\affiliation{Dipartimento di Fisica “E. R. Caianiello”, Università degli Studi di Salerno, via Giovanni Paolo II 132, I-84084, Fisciano, Salerno, Italy}
\author{C. Attanasio}
\affiliation{Dipartimento di Fisica “E. R. Caianiello”, Università degli Studi di Salerno, via Giovanni Paolo II 132, I-84084, Fisciano, Salerno, Italy}
\author{E. Scheer}
\affiliation{Department of Physics, University of Konstanz, 78457 Konstanz, Germany}
\author{A. Di Bernardo}
\affiliation{Department of Physics, University of Konstanz, 78457 Konstanz, Germany}
\affiliation{Dipartimento di Fisica “E. R. Caianiello”, Università degli Studi di Salerno, via Giovanni Paolo II 132, I-84084, Fisciano, Salerno, Italy}
\author{G. De Simoni}
\affiliation{NEST, Istituto Nanoscienze-CNR and Scuola Normale Superiore, Pisa, Italy}
\author{F. Giazotto}
 \affiliation{NEST, Istituto Nanoscienze-CNR and Scuola Normale Superiore, Pisa, Italy}

\email{Authors to whom correspondence should be addressed: sebastiano.battisti@sns.it, francesco.giazotto@sns.it}

 \title[]
  {Demonstration of high-impedance superconducting NbRe Dayem bridges}


\begin{abstract}
Here we demonstrate superconducting Dayem-bridge weak-links made of different stoichiometric compositions of NbRe. Our devices possess a relatively high critical temperature, normal-state resistance, and kinetic inductance. In particular, the high kinetic inductance makes this material a good alternative to more conventional niobium-based superconductors (e.g., NbN or NbTiN) for the realization of superinductors and high-quality factor resonators, whereas the high normal-state resistance yields a large output voltage in superconducting switches and logic elements realized upon this compound. Moreover, out-of-plane critical magnetic fields exceeding 2 T ensure that possible applications requiring high magnetic fields can also be envisaged. Altogether, these features make this material appealing for a number of applications in the framework of quantum technologies.
\end{abstract}

\maketitle

Rhenium is a superconductive material with a bulk critical temperature $T_c^{bulk}\sim2.4$K \cite{daunt1952}. When grown in thin films its superconductive properties slightly change and the critical temperature becomes $T_c^{film}\sim 1.85$K \cite{delsol2015} with a coherence length of $\xi \sim 150$nm \cite{ratter2017}. Rhenium compounds in general have attracted growing interest due to their unconventional superconductive properties \cite{kushwaha2024}. NbRe is a relatively-high critical temperature ($T_c^{bulk}\sim 9$K \cite{cirillo2015}) non-centrosymmetric superconductor \cite{shang2018}, which was proposed as a suitable material for superconducting single-photon detectors (SSPD) \cite{caputo2017,cirillo2020,cirillo2021,ejrnaes2022}. Thanks to a coherence length as short as $\xi\sim5$nm,  this material might provide a platform for the realization of ultra-thin superconducting films (thickness $t_{NbRe}\simeq 3.5$nm \cite{cirillo2016}) with large critical temperatures ($T_c\simeq 5.3$K) easily accessible in $^4$He closed-cycle refrigerators. Such a  short coherence length is the consequence of a sizable normal-state resistivity ($\rho_N \sim 3.0 x 10^{-5}$ $\Omega$m deduced by the normal state resistance and the dimensions of our structures), which turns out in a high $I_S R_N$ product, where $I_S$ is the switching current and $R_N$ is the normal-state resistance of the device, which is quite uncommon for full metallic systems due to their usually low resistances. This peculiar characteristic can be achieved with the use of dirty superconductors such as NbN. Differently, in our NbRe films this feature stems from their highly-disordered nature \cite{cirillo2016}.
Another crucial figure of merit of NbRe is its high kinetic inductance, which makes the devices based on this material appealing for all those applications requiring a high inductance such as superinductors \cite{bell2012,pop2014,manucharyan2009}, fluxonium qubits \cite{hazard2019}, high Q-factor resonators \cite{doyle2008}, radiation sensors \cite{kerman2007,santavicca2016}, superconductive oscillating circuits \cite{astafiev2012,schon2020}, and high inductance memory elements \cite{ilin2021}.
Here, we report on devices based on a Dayem-bridge (DB) geometry made of NbRe films with different stoichiometry. 
The DB structure can serve as a tool to realize several monolithic devices such as superconducting quantum interference devices (SQUIDs), magnetometers \cite{paoluccisquid}, and logic gates \cite{desimoni2020}. Furthermore, the opportunity to suppress the supercurrent of the DB via an external gate voltage \cite{desimoni2018} makes this kind of structure appealing for the realization of variable kinetic inductors with high dynamic range or nanosized superconducting switches with high output voltage due to the high $I_S R_N$ product ($\gtrsim 100$ mV) \cite{paolucci2019,desimoni2020,polini2022}.
\begin{figure}[t!]
  \includegraphics[width=1\columnwidth, trim= 2cm 3cm 0cm 4cm]{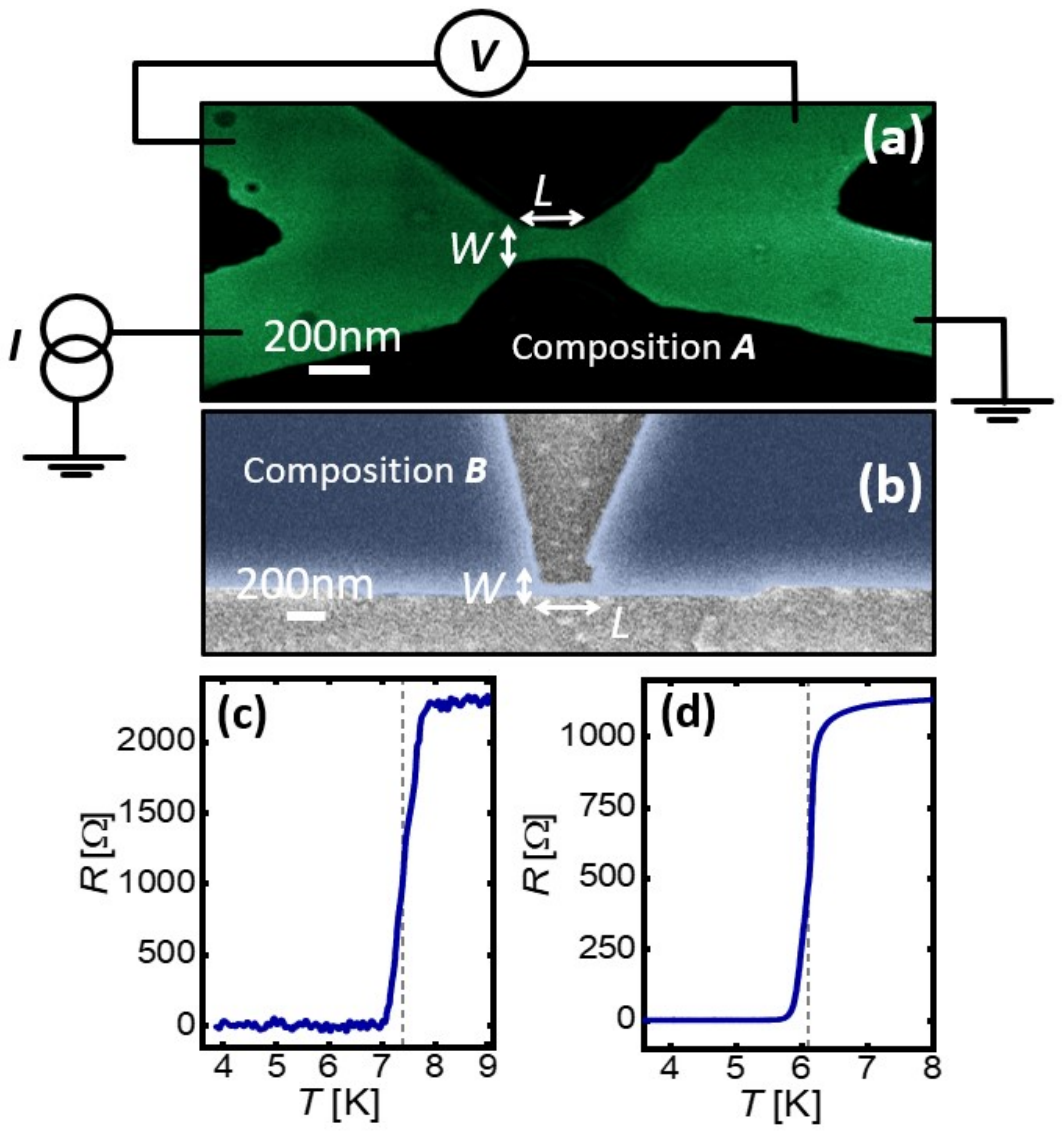}
    \caption{
    \textbf{(a)}: False color scanning electron micrograph (SEM) of one representative Dayem-bridge device with composition \textit{A}. The 4-wire measurement setup and the characteristic nanobridge  lateral dimensions are also indicated. $I$ indicates the current flowing through the device, while $V$ the voltage drop occurring across it.
    \textbf{(b)}: False color SEM image of a representative device with composition \textit{B} (the same 4-wire setup was used for this device). \textbf{(c)}: Resistance versus temperature curve ($R$ vs $T$) of the device shown in panel (a).  \textbf{(d)}: $R$ vs $T$ of the device shown in panel (b). The dashed lines indicate the critical temperatures $T_c^{exp}$ of the shown devices.}
  \label{fig1}
\end{figure}

In this work, we have investigated two different types of composition of NbRe thin films, i.e., composition \textit{A} and  \textit{B}, which have a stoichiometric ratio of Nb$_{0.1}$Re$_{0.9}$ and Nb$_{0.18}$Re$_{0.82}$, respectively.\\
We start by describing the fabrication process for the samples with composition \textit{A}. The Re-Nb films were prepared via magnetron sputtering using an ATC series UHV Hybrid deposition
system (AJA International, Inc.) with a base pressure of $1\times  10^{-8}$ Torr. The Re target (ACI Alloys, Inc.,
$99.99\%$ purity) and the Nb target (Testbourne, Ltd., $99.95\%$ purity) were both accommodated inside
$1.5"$ diameter DC guns. The sapphire substrate (AdValue Technology, thickness $650 \mu$m, C-cut) was cleaned
thoroughly with isopropyl alcohol before it was mounted on the holder. In the chamber’s configuration,
the substrate holder is at the center of the chamber facing upwards, while the sputtering guns are
located at the top. The substrate is rotated in plane throughout the whole deposition process to ensure
a homogeneous deposition layer over the whole surface. A pre-deposition in-situ cleaning of the
substrate involved heating it up to $700$ $^\circ$C for 10 minutes followed by a gentle bombardment of Ar$^+$ ions at
$600$ $^\circ$C for 5 minutes using a Kauffman source at $0.75$ mTorr. Plasma striking for guns was performed at $30$
mTorr with shutters closed and with the initial power $30$ W for each target. Then the gun powers were
ramped up to $150$ W for Nb and $500$ W for Re, with a decrease of the chamber pressure to $5$ mTorr in Ar flow.
Afterward, the deposition of Nb and Re took place. 
For better adhesion, the Re shutter was opened for $10$
seconds later than the Nb shutter. Simultaneous Nb and Re sputtering lasted $150$ s, after that the
Re shutter was closed, and in $10$ seconds the Nb shutter was also closed. After this, the temperature was again
raised to $900$ $^\circ$C for in-situ annealing during $30$ minutes and then cooled down to ambient temperature. All
the heating/cooling protocols consistently used a $30$ $^\circ$C/min ramp rate. 
To pattern the DB structures via reactive ion etching (RIE), a positive electron beam lithography (EBL) was performed with a PMMA mask on the sputtered films. Then $60$ nm of aluminum (Al) was deposited with an e-beam evaporator in order to have an Al hard mask, which is resistant to fluorine etching, on top of our NbRe film. The films were etched using a CF$_4$, Ar and O$_2$ gas mixture of 40/6/4 sccm at a power of $150$ W . Finally, the removal of the Al hard mask in an alkaline solution provides  the finished device shown in Fig.\ref{fig1}(a).\\
The composition \textit{B} films were sputtered via a 
stoichiometry NbRe (Nb$_{0.18}$Re$_{0.82}$) target ((Testbourne Ltd., 99.95\% purity)) with a power of 250 W and an Ar pressure of 3 mTorr at a base pressure of $1 \times 10^{-8}$ mTorr, which resulted in a deposition rate of $0.33$ nm/s. The 20-nm-thick NbRe films were sputtered on Al$_2$O$_3$ substrates (Crystec GmbH, thickness 430 $\mu$m, A-cut). To pattern the DBs via RIE, a negative lithography was performed with an etching-resistant negative resist. The etching was performed with an inductively-coupled plasma – 
reactive ion etcher (ICP-RIE), where the samples were etched for 30 s (at first 15 s and than three steps with 5 s after each checking the resistance in the wafer prober) in $26$ sccm Ar + $26$ sccm Cl$_2$  at $10$ mTorr with an RIE power of $20$ W and an ICP power of $750$ W (the sample plate was cooled to 10°C). A representative device with composition \textit{B}, produced with this process is shown in Fig.\ref{fig1}(b).\\
Both deposition methods have advantages and drawbacks. In fact the 
stoichiometry of the samples with composition A can be changed in order to change the film properties, which is clearly an advantage. However, the deposition method is more complicated and less reproducible with respect to the technique used for composition B samples. Moreover, the different etching processes used also present advantages and drawbacks: although the Al hard mask process with CF$_4$ RIE etching used for composition A samples, requires more fabrication steps, it relies on a positive resist lithography which can be more precise and the etching mask is removed very simply; the negative resist and Ar/Cl ICP RIE process used for composition B samples is faster but often presents issues regarding the negative resist mask removal after the etching process.\\


\begin{figure}[t!]
  \includegraphics[width=1\columnwidth,trim= 0cm 0cm -1cm 0cm]{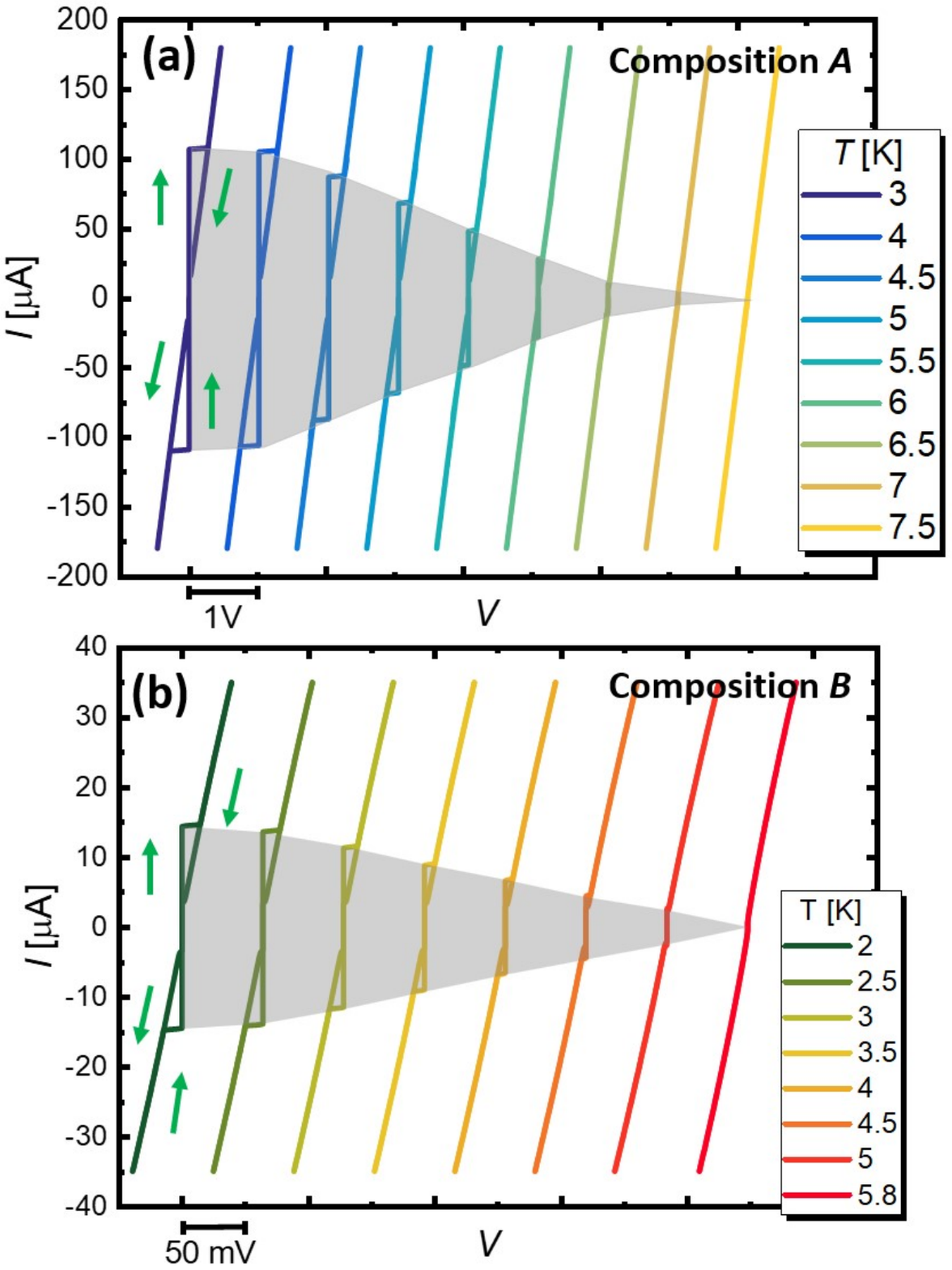}
    \caption{
    \textbf{(a)}[\textbf{(b)}]: Current vs voltage ($I-V$) characteristics for selected values of the bath temperature (in legend) for device 1A [(1B)]. The green arrows represent the current direction in the different branches of the curve. The gray areas correspond to the dissipationless region. The curves are horizontally shifted for clarity; each tick in the graph represents $1$V for panel \textbf{a} and $50$mV for panel \textbf{b}.}
  \label{fig2}
\end{figure}

\begin{figure}[t!]
  \includegraphics[width=0.9\columnwidth,trim= 2cm 2cm 1.5cm 1cm]{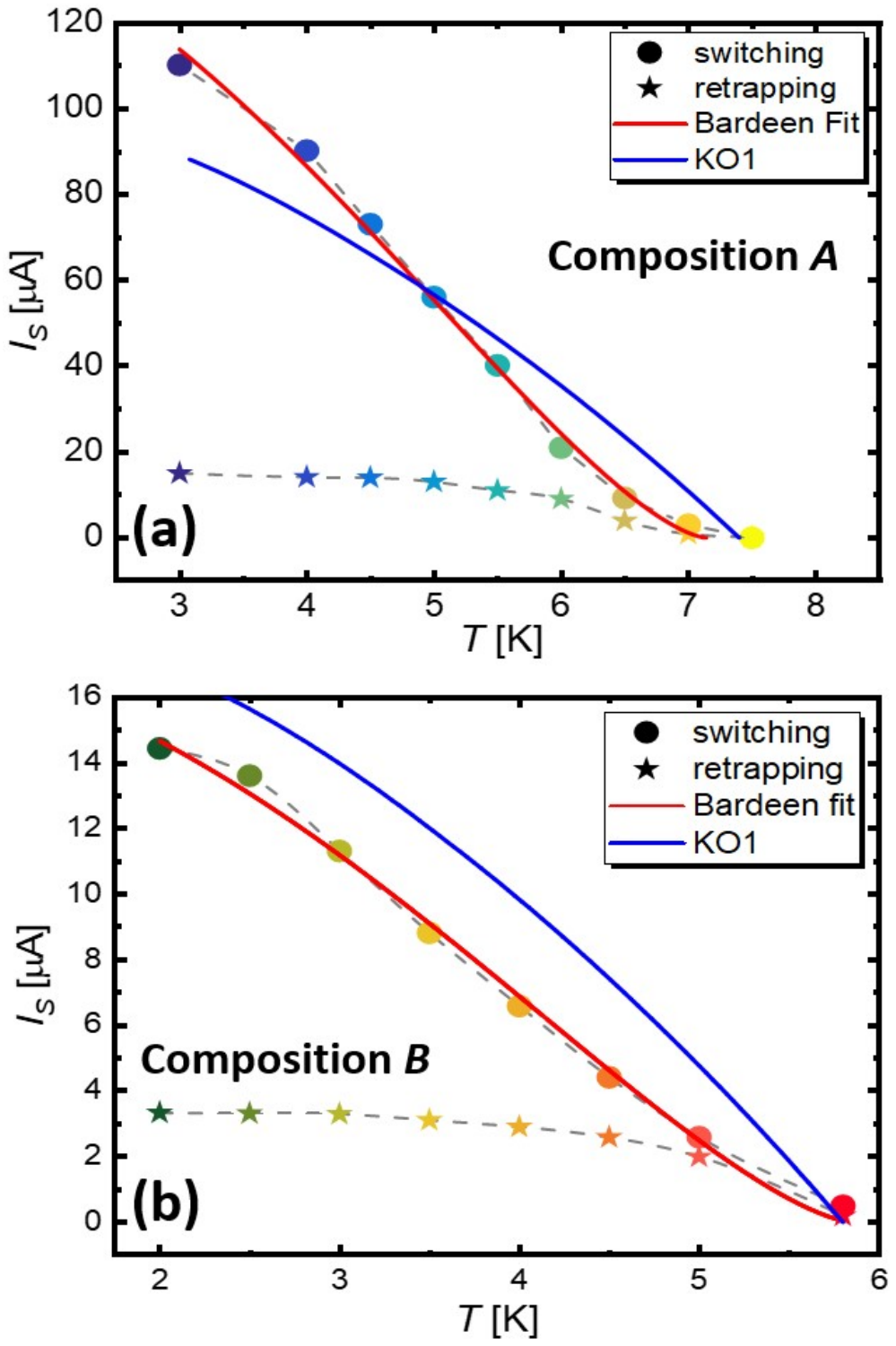}
    \caption{\textbf{(a)}[\textbf{(b)}]: Switching ($I_S$) and retrapping ($I_R$) currents as a function of bath temperature $T$ for a representative device with composition \textit{A}[\textit{B}]. The error bars are smaller than the size of the circles. The red line shows the accordance of a fit done with the Bardeen model in Eq.\ref{eq1}, whereas the green line shows the complete disagreement with the KO1 model in Eq. \ref{eq2a} calculated with the measured parameters. Dashed lines are guides to the eyes.}
  \label{fig3}
\end{figure}

\begin{table*}
\caption{The table summarises all the parameters of the measured devices. The geometrical features of the BDs: length $L$ and width $W$ were measured via SEM imaging, while the thickness $t$ was determined through deposition rate and time. $T_c^{exp}$ indicates the measured critical temperature, while $T_{c_{fit}}$ and $I_{S_{fit}}^0$ indicate respectively the estimated critical temperature and switching current obtained with a Bardeen fit as in Eq. \ref{eq1}. The product $I_S R_N$ is calculated for the maximum experimental switching current available, i.e.,  for the lower temperatures in Fig.\ref{fig2}(a-b). $R_{sq}$ and $L_K^{\square}$ indicate the sheet resistance and kinetic inductance respectively. $L_K^{\square}$ is taken at $T=4$K.}

\begin{ruledtabular}
\begin{tabular}{ccccc}

&\multicolumn{2}{c}{composition \textit{A}  (Re$_{0.9}$Nb$_{0.1}$)} &\multicolumn{2}{c}{composition \textit{B} 
 (Re$_{0.82}$Nb$_{0.18}$)} \\
 \hline
&Device 1\textit{A}&Device 2\textit{A}&Device 1\textit{B}
&Device 2\textit{B}\\
\hline
 \textit{L} [nm] &180 &170 &200 &200  \\
 \textit{W} [nm] &80 &75 &64 &56  \\
 \textit{t} [nm] &30 &30 &20 &20  \\
 $T_c^{exp}$ [K]&$7.4\pm 0.1$&$7.4\pm 0.1$ &$6.1\pm0.1$ &$6.1\pm0.1$  \\
 $T_{c_{fit}}$ [K]&$7.1\pm 0.1$&$7.0\pm0.1$ &$5.82\pm0.03$ &$5.91\pm0.06$  \\
 $I_{S_{fit}}^0$ [$\mu$A]&$158\pm 7$ &$136\pm2$ &$33\pm3$ &$18\pm3$  \\
 $R_N$ [$\Omega$]&$2276\pm 49$ &$2301\pm38$ &$1136\pm5$ &$1171\pm6$  \\
 $I_S R_N$ [mV]&$250\pm5$ & $232\pm4$ &$24.4\pm0.4$ & $14.4\pm0.3$ \\
 $\rho$ [$\Omega$m]&$(3.0 \pm 0.6) 10^{-5}$&$(3.0 \pm 0.6) 10^{-5}$ &$(7.3\pm0.2)10^{-6}$ &$(6.6\pm0.2)10^{-6}$\\
 $R_{sq}$ [$\Omega$]&$1011\pm24$ &$1015\pm17$& $363\pm2$&$328\pm2$\\
 $L_K^{\square}$ [pH]&$135\pm3$&$137\pm3$&$85\pm1$&$77\pm1$
 \label{tab}
\end{tabular}
\end{ruledtabular}
\end{table*}

The devices were then wired, as displayed in Fig\ref{fig1}(a), with a 4-wire standard setup in order to take the full characterization of their properties. The samples were cooled down in $^4$He cryogen-free cryostats with a base temperature of $\ 3$K. In Fig.\ref{fig1}(c) the characteristic $R$ vs $T$ shows that the critical temperature of the devices with composition \textit{A} is $T_{c_{1A}}^{exp}\sim 7.4$K, while for composition \textit{B} is $T_{c_{1B}}^{exp}\sim 6.1$K [see Fig.\ref{fig1}(d)]. The $T_c^{exp}$ value for both samples is taken when $dR/dT$ is maximum [dashed lines in Figs.\ref{fig1}(c,d)]. We note that both devices have only one visible transition in the curve shown in Fig.\ref{fig1}(c),(d): this is because the coherence length ($\xi\sim5$nm) of this material is much lower than the lateral dimensions of the DB (see Tab.\ref{tab}), resulting in a negligible change of $\Delta (T)$ in the constriction.\\
Figure \ref{fig2}(a) shows the current-voltage (\textit{I-V}) curves of device 1\textit{A} for different temperatures horizontally shifted for clarity. It can be seen that at $T_{bath}=3$K the switching current is $I_S=107 \mu$A and then rising the temperature it lowers monotonically, becoming vanishing small at $T\sim 7.4$K. Figure \ref{fig2}(a) marks also the hysteretical behavior of these DBs, stemming from the Joule heating occurring in the nanobridge when in the normal state. 
Figure \ref{fig2}(b) reports the same characterization for the device 1\textit{B}, which displays lower switching (and retrapping) currents, and critical temperature. In fact, we have that for $T_{bath}=3$K the switching current is $I_S=11 \mu$A. With respect to device 1\textit{A}, device 1\textit{B} is narrower and thinner, resulting in both a weaker superconductivity (because of the reduction of the superconductive energy gap $\Delta$ when shrinking the dimensions \cite{tinkham2004}) and in a smaller section of the superconductor, which, given a critical current density, causes the reduction of the critical current itself. In fact, the  critical temperature of device 1\textit{B}, $T_{c_{1B}}\simeq 6.1$K stems for a weaker superconductivity with respect to composition \textit{A} devices. An analogous discussion can be done for the other devices investigated in this study (see Tab.\ref{tab}). However, these differences in the geometrical properties of the device can only partially address this discrepancy. In fact it must also be noticed that a different Re percentage in the films changes its superconducting properties, in particular, the normal-state resistance. From the curves shown in Fig.\ref{fig1}(c,d) the device's normal-state resistance $R_N$ can be extracted. 
This parameter is particularly relevant for applications of these structures since it determines the output voltage ($I_S R_N$) of the device when it switches from the superconducting to the normal state. This number can determine whether the output signal of the switching device can be easily detected or even reused to control other structures. With conventional metallic systems such as aluminum, titanium or niobium this product can reach values of $\sim 20$ mV \cite{desimoni2020}. For our devices the measured values are collected in Tab.\ref{tab}, and one can observe that this value is an order of magnitude higher. Moreover it can be seen that $R_N^{A}\simeq 2R_N^{B}$. Taking into account the nanobridge's geometrical parameters, we argue that this behavior might originate from the higher percentage of Re in the composition of \textit{A} devices. For this reason, we believe that concerning applications that require a high $I_S R_N$ product ($\gtrsim 100$ mV), such as the recently investigated superconducting gate-tunable transistors \cite{desimoni2018,dibernardo2023,paolucci2019} or other superconducting switching devices such as cryotrons \cite{buck1956}, nanocrytrons \cite{mccaughan2014} or more complex structures such as shift registers \cite{buzzi2023} or memories \cite{foster2023} built with switching devices as basic elements, the composition \textit{A} promises better performances. 
Figure \ref{fig3} shows the trend of the switching current $I_S$ and the retrapping current $I_R$ as a function of bath temperature $T$ for both samples 1\textit{A} and 1\textit{B}. The curves were fitted with the Bardeen equation \cite{bardeen1962} that holds for a thin superconducting wire:
\begin{equation}
I_S(T)=I_{S_{fit}}^0 \left[1-
 \left(\frac{T}{T_{c_{fit}}}\right)^2%
\right]^{3/2},
\label{eq1}
\end{equation}
where $I_{S_{fit}}^0$ and $T_{c_{fit}}$ are the fitting parameters reported in Tab.\ref{tab}.
Furthermore, we  verified the agreement with the Kulik-Omelyanchouk model in the diffusive regime (dirty limit) KO1 \cite{golubov2004}, 
\begin{subequations}
\begin{equation}
I_S(T)= \text{max}_{\phi} (I (\phi)),
\label{eq2a}
\end{equation}
\begin{equation}
I_S(\phi)=\frac{4\pi k_B T}{e R_N} \sum_{\omega>0}\frac{\Delta(T) \text{cos}(\phi/2)}{\delta}\text{arctan}[\frac{\Delta(T) \text{sin}(\phi/2)}{\delta}] ,
\label{eq2b}
\end{equation}
\end{subequations}
calculated with the measured parameters for $R_N$ and $T_c$. In Eq.\ref{eq2b}, 
$\delta=\sqrt{\Delta(T)^2\text{cos}^2(\phi/2)+\omega^2}$, $\omega=\pi k_BT(2n+1)$ are the Matsubara frequencies, $k_B$ is the Boltzmann constant, $\phi$ is the phase difference across the weak-link, and $\Delta(T)=1.764k_B T_c^{exp}\textrm{tanh}[1.74\sqrt{T_c^{exp}/T-1}]$ is the formula used for the temperature dependence of the superconductive gap.
It is worth noting that it was not possible to obtain a reasonable agreement between the KO1 model and the experimental data, thereby confirming that the length of the wire is much greater than the superconducting coherence length, i.e., $L\gg\xi$. We note, indeed, that the KO1 model holds for  $L\ll\xi$ \cite{golubov2004} and here it was used to demonstrate the \textit{non-strictly-Josephson} nature of these DBs. On the other hand, the Bardeen formula fits well the experimental data over the full temperature range, yielding a coefficient of determination ${\cal R}^2 > 0.99$. This highlights the BCS nature of the NbRe thin films.

\begin{figure}[t!]
  \includegraphics[width=1\columnwidth,trim= 1cm 0cm 0cm 0cm]{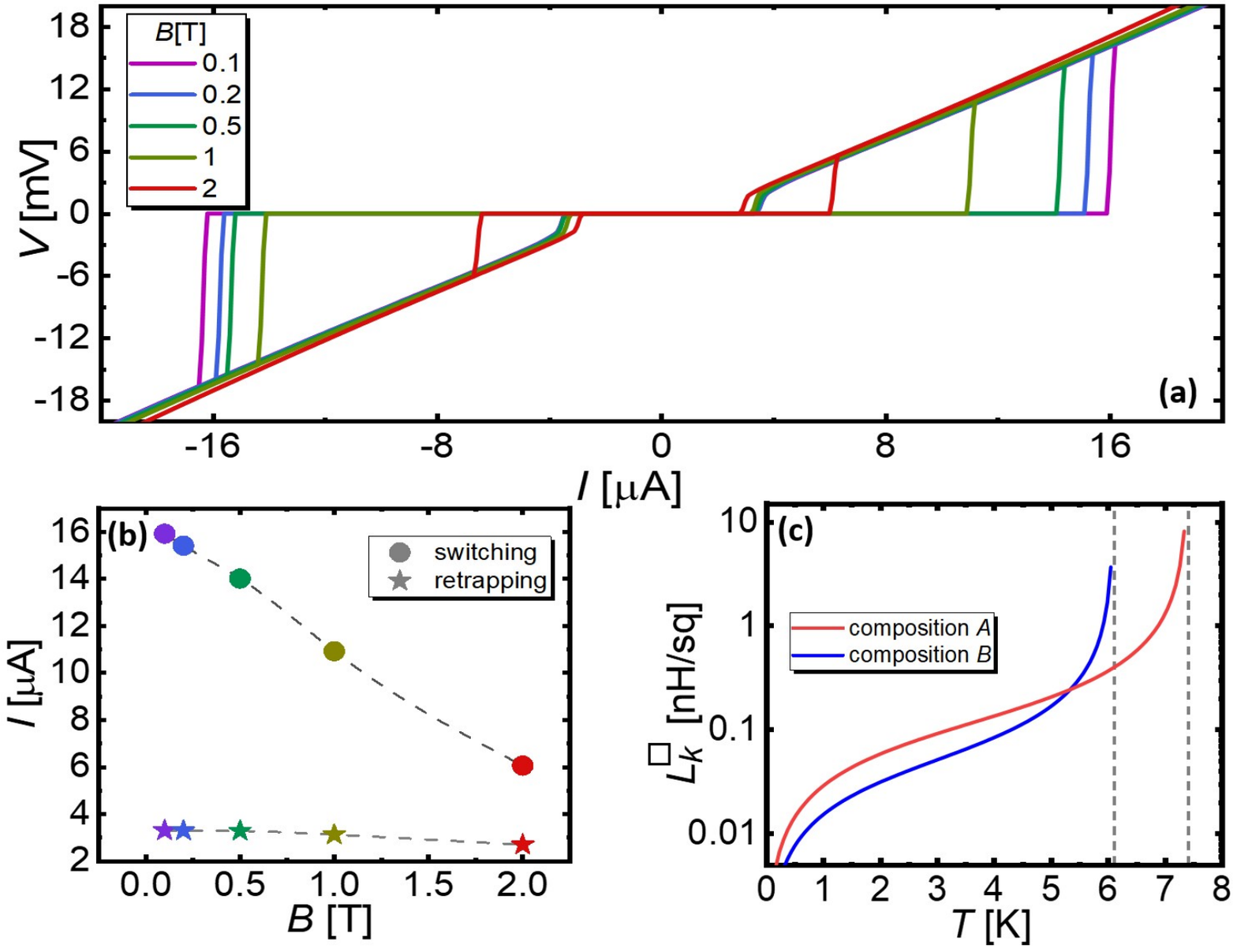}
    \caption{\textbf{(a)} $I-V$ characteristics as a function of perpendicular-to-plane magnetic field $B$ applied to a representative device with composition \textit{B}. \textbf{(b)}: Switching and retrapping currents of the curves in (a) as a function of $B$. Dashed lines are guides to the eye. \textbf{(c)}: Kinetic inductance per square $L_K^\square$ as a function of temperature $T$ for samples 1A and 1B. Dashed lines indicate the asymptotes of $L_K^\square$ due to $\Delta (T)$ approaching $0$ at $T=T_c$. All the measurements shown in panels (a) and (b) were taken at $T_{bath}=2K$.
    } 
  \label{fig4}
\end{figure}

Figure \ref{fig4}(a) shows several \textit{I-V} curves measured at 2K for increasing values of the external perpendicular-to-plane magnetic field for a device with composition \textit{B}. The full trend of the switching and retrapping currents as a function of the magnetic field is highlighed in Fig.\ref{fig4}(b). Up to $2$T (i.e., the maximum magnetic field that our magnet could support), the NbRe DB still shows a sizable switching current of the order of $\sim 6\mu$A. 
This suggests that NbRe has a perpendicular-to-plane critical magnetic field at least comparable to other high-kinetic inductance superconductors \cite{suzuki1987,kakhaki2023}. We emphasize that the robustness of this material to magnetic fields makes it an optimal candidate for experimental situations and applications where high magnetic fluxes cannot be avoided or are even required such as \textit{e. g.}, quantum Hall experiments in hybrid platforms \cite{rickhaus2012}.

Figure \ref{fig4}(c) shows the kinetic inductance per square ($L_K^\square$) calculated within the BCS theory for a superconducting stripe (indeed in our devices we have $W > 10 \xi$) \cite{annunziata2010}
\begin{equation}
    L_K^\square (T)=\frac{R_{sq} \hbar}{\pi \Delta (T)}\frac{1}{\textrm{tanh}(\Delta (T)/k_B T)},
    \label{eq3}
\end{equation}
using for $R_{sq}$ and $T_c$ the experimentally measured values  reported in Tab.\ref{tab}. We stress that, in principle,  $L_K^\square$ obtains values as large as several hundreds of pH/$\square$ at a few Kelvins, which is comparable or larger than that obtained in TiN $\simeq 40$pH \cite{joshi2022}, NbN $\simeq 82$pH/$\square$ \cite{niepce2019} and NbTiN $\simeq 115$pH/$\square$ \cite{harvey2020} thin films for $T>2$K.
For $T\gtrsim 4$K, the device with composition \textit{B} has a higher $L_K^\square$ with respect to that with composition \textit{A}. This might be due to the lower $T_c$ and $\Delta (T)$ of \textit{B}-devices. Indeed, although $L_K^\square$ is proportional to$R_{sq}$, Eq. \ref{eq3} reminds us that the kinetic inductance is also exponentially dependent on $\Delta(T)$. However, it is worth highlighting that by reducing the DB cross-section, in order to intentionally lower the critical temperature $T_c$ and the gap $\Delta (T)$ of devices with composition \textit{A}, it would be possible to take advantage of the larger resistivity of the films with higher Re concentration to achieve improved performances in terms of kinetic inductance.\\
In general, we can conclude that by varying the Re percentage of these films one can explore the parameters space and find the better performances in terms of $L_K^{\square}$ and $I_S R_N$ for a given geometry.

In summary, we have reported the fabrication and characterization of NbRe superconducting Dayem nanobridges. The better performance, in terms of $T_c$, $L_K$, and $I_S R_N$, of the devices with composition \textit{A} suggests that a higher percentage of Re is beneficial for applications requiring these characteristics. The achievable values of the kinetic inductance, which are higher than the ones reported in literature for NbN \cite{niepce2019}, suggest that the NbRe can be used in several applications requiring high $L_K$ such as superinductors \cite{bell2012,pop2014,manucharyan2009}, radiation sensors \cite{caputo2017,cirillo2020,cirillo2021}, and high-quality factor resonators \cite{doyle2008}. Moreover, the sizable critical magnetic field highlights its possible exploitation in qubit systems \cite{hazard2019}. 
Given all these features, we believe that this superconducting metal might be considered as a promising alternative to more conventional Nb compounds (e.g., NbN and NbTiN) for quantum technology applications \cite{polini2022}.

\begin{acknowledgements}We acknowledge the EU’s Horizon 2020 Research
and Innovation Framework Programme under Grant
No. 964398 (SUPERGATE), No. 101057977 (SPECTRUM), and the PNRR MUR project PE0000023-
NQSTI for partial financial support. We also acknowledge partial support of this research by US ONR grants No. N00014-21-1-2879,  No. N00014-20-1-2442 and No. N00014-23-1-2866.
\end{acknowledgements}

\bibliography{ReNb}

\end{document}